\documentclass[10pt]{iopart}
\usepackage{graphicx}  
\begin{document}
\vspace*{-3em}
\title{Kaon properties and cross sections in nuclear medium} 

\author{K. Tsushima\dag, A. Sibirtsev\ddag\ and A.W. 
Thomas\dag\footnote[3]{ADP-00-42/T425\\
Presented by K. T. at the 5th International
Conference on Strangeness in Quark Matter, July 20 -- 25, 2000, 
Berkeley, California. \\ 
Supported by the
Australian Research Council and the Forschungszentrum J\"ulich.}}

\address{\dag\ Special Research Center for the Subatomic Structure of Matter 
and Department of Physics and Mathematical Physics, 
University of Adelaide, SA 5005, Australia}

\address{\ddag\ Institut f\"ur Theoretische Physik, Universit\"at Giessen,
D-35392 Giessen, Germany}

\begin{abstract}
Results for the $\pi + N \to  \Lambda, \Sigma + K$ reactions 
in nuclear matter of Ref.~\cite{Tsushima0}  
are presented. To evaluate the in-medium
modification of the reaction amplitude as a function of the
baryonic density we introduce relativistic, mean-field potentials
for the initial, final
and intermediate mesonic and baryonic states in the resonance model. 
These  vector and
scalar potentials were calculated using the quark meson coupling model.
Contrary to earlier work which has not allowed for the change of the
cross section in medium, we find that the data for kaon production
at SIS energies are consistent with a repulsive 
$K^+$-nucleus potential.
\end{abstract}

%
%

\vspace{-1.5em}
\section{Introduction}
The properties of kaons in nuclear matter have recently  attracted
enormous interest because of their capacity to signal chiral symmetry
restoration or give information on the possibility of kaon
condensation in neutron
stars~\cite{Kaplan}. Studies with a variety of
models~\cite{Waas1,Tsushima1} indicate that the
antikaon potential is attractive while the kaons feel a repulsive
potential in nuclear matter. 
However, the analysis of  available data on
$K^+$ production from heavy ion collisions at SIS
energies~\cite{Barth,Senger3} contradicts
the predictions that the kaon potential is repulsive.
The comparison between the heavy ion calculations and the
data~\cite{Senger3,Li2,Bratkovskaya} indicates
that the $K^+$-meson spectra are best described by neglecting
any in-medium modification of the kaon properties. Furthermore, the
introduction of even a weakly repulsive
$K^+$-nucleus potential results in a
substantial underestimate of the experimental data on kaon
production.

Since in  heavy ion collisions at
SIS energies~\cite{Barth,Schroter}
the $K^+$-mesons are predominantly produced by secondary pions,
we investigate the kaon production reactions,
$\pi + N \to \Lambda, \Sigma + K$,
in nuclear matter. We calculate the in-medium
reaction amplitudes, taking into account the scalar and vector
potentials for incident, final and intermediate mesons and baryons.

\section{Mean-field potentials for mesons and baryons}
We use the quark-meson coupling (QMC) model~\cite{Guichon},
which has been successfully applied to the various problems
of nuclear physics~\cite{Guichonf}
and the studies of meson and hyperon properties in a nuclear
medium~\cite{Tsushima1,Tsushima2}.
The Dirac equations for the quarks and antiquarks in the hadron bags 
($q = u,\bar{u},d$ or $\bar{d}$, hereafter), 
neglecting the Coulomb force,
are given by~\cite{Tsushima1}:
\begin{eqnarray}
\hspace*{-7em}
\left[ i \gamma \cdot \partial_x -
(m_q - V^q_\sigma)
\mp \gamma^0
\left( V^q_\omega +
\frac{1}{2} V^q_\rho
\right) \right]
\left( \begin{array}{c} \psi_u(x)  \\
\psi_{\bar{u}}(x) \\ \end{array} \right)
&=& 0,\,\,
\left( -\frac{1}{2}V^q_\rho\, {\rm for}
\left(\begin{array}{c} \psi_d  \\
\psi_{\bar{d}}\\ \end{array} \right) \right),
\label{diracu}
\\
\left[ i \gamma \cdot \partial_x - m_{s} \right]
\psi_{s} (x)\,\, ({\rm or}\,\, \psi_{\bar{s}}(x)) &=& 0.
\label{diracs}
\end{eqnarray}
The mean-field potentials for a bag in  nuclear matter
are defined by $V^q_\sigma = g^q_\sigma
\sigma$,
$V^q_\omega = $ $g^q_\omega
\omega$ and
$V^q_\rho = g^q_\rho b$,
with $g^q_\sigma$, $g^q_\omega$ and
$g^q_\rho$ the corresponding quark-meson coupling
constants. 

The hadron masses
in symmetric nuclear matter relevant for the present study 
are calculated by:
\begin{equation}
\hspace*{-6em}
m_h^* = \frac{(n_q + n_{\bar{q}}) \Omega_q^*
+ (n_s + n_{\bar{s}}) \Omega_s - z_h}{R_h^*}
+ {4\over 3}\pi R_h^{* 3} B,\hspace{1em}
{\rm with}\hspace{1em} \left. \frac{\partial m_h^*}
{\partial R_h}\right|_{R_h = R_h^*} = 0,
\label{equil}
\end{equation}
where $\Omega_q^*
 = \sqrt{x_q^2{+}(R_h^* m_q^*)^2}$, with
$m_q^* = m_q{-}g^q_\sigma \sigma$ and
$\Omega_{s} = \sqrt{x_{s}^2{+}(R_h^* m_{s})^2}$, and 
$x_q$ and $x_{s}$ being the bag eigenfrequencies of the corresponding 
quarks. 
In Eq.~(\ref{equil}), $n_q$ ($n_{\bar{q}}$) and $n_s$ ($n_{\bar{s}}$)
are the lowest mode light quark (antiquark) and strange 
(antistrange) quark numbers in the hadron, $h$, respectively,
and the $z_h$ parametrize the sum of the
center-of-mass and gluon fluctuation effects, and are assumed to be
independent of density. The parameters are determined in free space to
reproduce their physical masses.
We chose the values $m_q$ = 5 MeV and 
$m_s$ = 250 MeV for the current quark masses, and $R_N = 0.8$~fm 
for the bag radius of the nucleon in free space~\cite{Tsushima1}. 
We stress that 
only three coupling constants, $g^q_\sigma$, $g^q_\omega$
and $g^q_\rho$, are adjusted to fit nuclear data -- namely the
saturation energy and density of symmetric nuclear matter and the bulk
symmetry energy. 
Exactly the same coupling constants, $g^q_\sigma$, $g^q_\omega$ and
$g^q_\rho$, are used for the light quarks in the mesons and hyperons as in 
the nucleon. 
However, for the kaon system it was
phenomenologically necessary to increase the strength of the vector
coupling to the non-strange quarks in the $K^+$,   
and thus, we will use the stronger vector potential,
$1.4^2 V^q_\omega$, for the $K^+$-meson~\cite{Tsushima1}.   
The scalar ($U^{h}_s$) and vector ($U^{h}_v$) potentials 
felt by the hadrons, $h$,  
in nuclear matter are self-consistently calculated by:
\begin{equation}
\hspace*{-6em}
U^{h}_s
\equiv U_s = m^*_h - m_h,
\hspace{1em}
U^{h}_v =
  (n_q - n_{\bar{q}}) {V}^q_\omega - I_3 V^q_\rho, 
\,\,(V^q_\omega \to 1.4^2 {V}^q_\omega\,\, {\rm for}\, K^+), 
\label{vdpot}
\end{equation}
where, $I_3$ is the third component of isospin projection  
of the hadron, $h$, and the $\rho$ meson mean field potential, $V^q_\rho$,
is zero in symmetric nuclear matter.
Then, within the approximation that the mean field potentials are 
independent of momentum, 
the four-momentum of the hadron is modified by, 
$p^\mu_h = (\sqrt{{\mbox{\boldmath $p$}}^2 + m_h^{*2}} + U^{h}_v, 
{\mbox{\boldmath $p$}}$), which 
modifies not only the kinematical factors such as the flux,  
the phase space and the threshold, 
but also modifies the reaction amplitudes.

Furthermore, we include the in-medium modification of 
the resonance masses, which appear in the reaction amplitudes
in the resonance model.
In view of its numerous successful applications elsewhere, we base
our estimate on the QMC model~\cite{Tsushima1,Tsushima2}, and use 
those estimated values~\cite{Tsushima0} for the in-medium resonance masses.

\section{$\pi + N \to \Lambda + K$ in nuclear matter}

\begin{figure}[htb]
\vspace{-6em}
\begin{center}
\includegraphics[width=10cm,height=11cm]{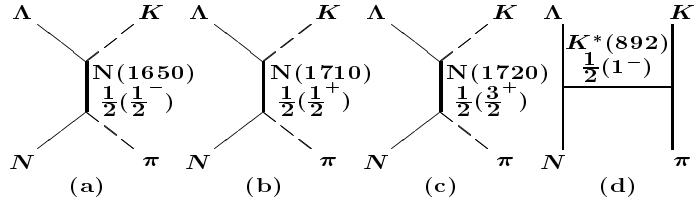}
\end{center}
\vspace{-19em}
\caption{\label{pbyklafig}
$K$ and $\Lambda$ production processes included in 
the resonance model~\protect\cite{Tsushima3,Tsushima1}.
\vspace{-1.5em}
}
\end{figure}

Now we apply the resonance model~\cite{Tsushima3,Tsushima1} 
to calculate the in-medium
amplitudes focusing on $\pi + N \to \Lambda + K$.
We consider kaon and hyperon production processes in $\pi N$ collisions
shown in Fig.~\ref{pbyklafig}. 
(See Refs.~\cite{Tsushima3,Tsushima0} for $\pi + N \to \Sigma + K$.) 

We extend the model by including medium modification
of the hadron properties, not only in the kinematic factors such as
the flux and the phase space, but also in the reaction amplitudes, 
as discussed in the previous section.
(See also Ref.~\cite{Tsushima0}.)

We found that the maximal downward shift of the
reaction threshold in nuclear matter occurs at
baryon densities around $\rho_B \simeq 0.6 \rho_0$ 
($\rho_0$ = 0.15~fm$^{-3}$), and 
the maximum of the downward shift of the
$\pi + N \to \Lambda, \Sigma + K$ reaction threshold amounts to roughly
30~MeV. We also found that at  baryon densities
$\rho_B >$ 0.2~fm$^{-3}$ thresholds for the 
$\pi + N \to \Lambda, \Sigma + K$ reactions 
are higher than those in free space.

\begin{figure}[htb]
\begin{center}
\vspace{-1.5em}
\includegraphics[width=8.0cm,height=5.5cm]{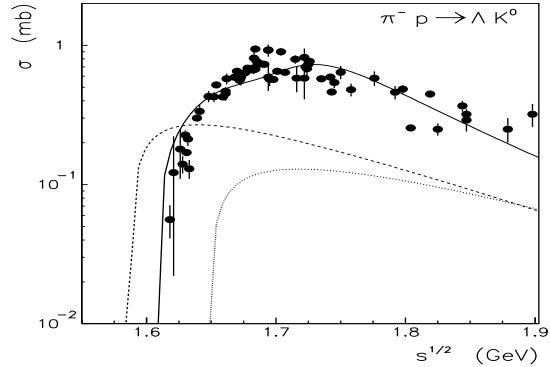}
\end{center}
\vspace{-1.5em}
\caption{\label{laka4}Energy dependence of the 
total cross section, $\pi^- + p \to \Lambda + K^0$, 
as a function of the invariant collision energy, $\sqrt{s}$, 
calculated for different baryon densities.
The data in free space are taken from Ref.~\protect\cite{LB}. 
The lines indicate our results for
free space (solid) and for nuclear matter at baryon density 
$\rho_B = \rho_0$ (dashed) and $\rho_B = 2 \rho_0$ (dotted) 
($\rho_0$ = 0.15~fm$^{-3}$).}
\vspace{-1.5em}
\end{figure}

We show the energy dependence 
of the total $\pi^- + p \to \Lambda + K^0$
cross section in Fig.~\ref{laka4}.
The calculations for free space 
are in reasonable agreement with the data, as shown by the solid line.
The dashed line in Fig.~\ref{laka4} shows the results obtained for  
nuclear matter at $\rho_B = \rho_0$, while the
dotted line is the calculation at $\rho_B = 2 \rho_0$.

\section{Impact on heavy ion studies}
It is expected that in relativistic heavy ion 
collisions at SIS energies  nuclear matter can be compressed up to 
baryonic densities of order $\rho_B \simeq 3 \rho_0$~\cite{Senger3}. 
The calculation of the time
and spatial dependence of the baryon density distribution is
a vital aspect of  dynamical heavy
ion simulations. 
The baryon density
$\rho_B$ available in heavy ion collisions evolves with the interaction 
time, $t$, and is given by the dynamics of the heavy ion collision.
In the following estimates we investigate
the density dependence of the production cross section for central 
central heavy ion collisions. 

To calculate the $K^+$-meson production cross section averaged
over the available density distribution we adopt the 
density profile function obtained by dynamical 
simulations~\cite{Hombach} for $Au + Au$ collisions at 
2 AGeV and at impact 
parameter $b = 0$. This can be parametrized as 
\begin{equation}
\label{time1}
\rho_B (t) = \rho_{max} \, \exp 
\left( [\, t\,- \,{\bar t} \, \, ]^2/{\Delta t}^2
\right),
\label{dprofile}
\end{equation} 
where the parameters, $\rho_{max}$ = 3 $\rho_0$, 
${\bar t}$ = 13~fm and ${\Delta t}$ = 6.7~fm, were fitted to the heavy
ion calculations~\cite{Hombach}.

\begin{figure}[htb]
\begin{center}
\vspace{-1.5em}
\includegraphics[width=8.0cm,height=5.5cm]{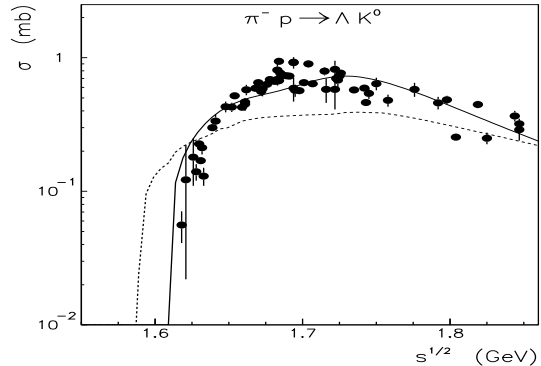}
\vspace{-1.5em}
\caption{\label{laka4a}Energy dependence of the
total cross section, $\pi^- + p \to \Lambda + K^0$, 
as a function of the invariant collision energy, $\sqrt{s}$.
The data in free space are taken from Ref.~\cite{LB}. The 
solid line indicates our calculation for
free space. The dashed line shows the cross section calculated 
by averaging over the density function profile~\protect\cite{Hombach}
given by the time evolution obtained for 
$Au + Au$ collisions at 2 AGeV (see Eq.~(\protect\ref{dprofile})).}
\vspace{-1.5em}
\end{center}
\end{figure}

The total cross section for the $\pi^- + p \to \Lambda + K^0$ reaction 
integrated over the time range $5 \le t \le 23$~fm and weighted by 
the time dependent density profile given in Eq.~(\ref{dprofile}),  
is shown by 
the dashed line in Fig.~\ref{laka4a}. The 
limits of the $t$ integration were taken from the simulations of the
$Au + Au$ collision time evolution in Ref.~\cite{Hombach}. The circles and 
solid line in Fig.~\ref{laka4a} show the experimental data 
in free space~\cite{LB}
and the calculations in free space, respectively. 

One can see that the total cross section averaged over 
the collision time (time dependent density profile) for 
the $\pi^- + p \to \Lambda + K^0$ reaction 
is quite close to the result given in
free space integrated  
up to $\sqrt{s} \simeq$ 1.7 GeV.
Indeed, the results shown in Fig.~\ref{laka4a} actually explain why the 
heavy ion calculations with the free space kaon production cross
section might quite reasonably reproduce the experimental data. 

Furthermore, the total cross section averaged over 
the time dependent density profile,  
shown by the dashed line in Fig.~\ref{laka4a}, 
should additionally be averaged over the
invariant collision energy distribution available in heavy
ion reactions. The number of meson-baryon collisions, $N_{mB}$, 
for the central $Au + Au$ collisions at 2 AGeV is given 
in Ref.~\cite{CBJ} as a function of the invariant collision energy, 
$\sqrt{s}$. It can be parametrized for $\sqrt{s} > $1~GeV as
\begin{equation}
dN_{mB} / d\sqrt{s} = N_0 \, \exp{\left( [\, 
\sqrt{s} \,- \,{\sqrt{s_0}} \, \, ]^2/[{\Delta \sqrt{s}} ]^2
\right)},
\label{donsi}
\end{equation}
where the normalization factor $N_0$ = 6$\times$10$^4$ GeV$^{-1}$, 
while $\sqrt{s_0}$ = 1 GeV
and $\Delta \sqrt{s}$ = 0.63~GeV. Note that, at SIS
energies $N_{mB}$ is almost entirely given by the pion-nucleon
interactions, and heavy meson and baryon collisions contribute 
only to the high energy tail of the 
distribution in Eq.~(\ref{donsi}) -- with quite small densities~\cite{CBJ}.  
{}Finally, if we also average the 
calculated, in-medium, total cross section for 
$\pi^- + p \to \Lambda + K^0$, shown by the dashed line
in Fig.~\ref{laka4a}, over the available energy distribution given
in Eq.~(\ref{donsi}), we obtain an average total 
kaon production cross section of 
${<}K{>}$ = 65~$\mu$b for 
central $Au + Au$ collisions at 2 AGeV. 
This result is  indeed 
compatible with the calculations using the free space total cross section 
of the $\pi^- + p \to \Lambda + K^0$ reaction, 
which provide an average total kaon production 
cross section of ${<}K{>}$ = 71~$\mu$b
for central $Au + Au$ collisions at 2 AGeV. 
Note that the inclusion of even a slight modification of the 
$K^+$ mass because of the nuclear medium 
(without the corresponding changes introduced here) 
leads to a substantial reduction of the inclusive $K^+$ spectra
(by as much as a factor of 2 or 3), 
compared to that calculated using the free space properties 
for the relevant hadrons~\cite{Bratkovskaya}.


We stress that at SIS energies  reaction channels with 
a $\Sigma$-hyperon in the final state play a minor role, 
because of the upper limit of the energy available in the 
collisions.  
The downwardly shifted $\pi + N \to \Sigma + K$ reaction threshold 
at small densities is still quite high.  

\section{Conclusion}
Our present study shows that if one accounts for 
the in-medium modification of the production amplitude   
(i.e., the in-medium properties of the $K^+$-meson and hadrons) correctly,  
it is possible to understand $K^+$ production data in heavy ion collisions 
at SIS energies, even if the $K^+$-meson feels the theoretically expected,  
repulsive mean field potential. The apparent failure to explain  
the $K^+$ production data if one includes the purely kinematic 
effects of the in-medium modification 
of the $K^+$-meson and hadrons, appears to be a consequence of the
omission of these effects on the reaction amplitudes.


\vspace{2ex}
\noindent
{\bf References}
\vspace{1ex}
\begin{thebibliography} {99}
\bibitem{Tsushima0}
        Tsushima K, Sibirtsev A, Thomas A W, {\it nucl-th/0004011}.
\bibitem{Kaplan}
        Kaplan D B and Nelson A E 1986 {\it Phys. Lett. B} 
        {\bf 175} 57; (E) 1986 {\it ibid.} {\bf 179} 409;
        Brown G E and Rho M 1991 {\it Phys. Rev. Lett.} {\bf 66} 2720;
        Ko C M et al. 1991 {\it Phys. Rev. Lett.} {\bf 66} 2577;  
        1991 {\it Phys. Rev. Lett.} {\bf 66} 2577;  
        1991 {\it ibid.} {\bf 67} 1811;  Lee C H et al. 
        1995 {\it Nucl. Phys. A} {\bf 585} 401;
        Glendenning N et al. 2000 {\it Phys. Rev. C}
        {\bf 62} 025804.
\bibitem{Waas1}
        Waas T et al.  1997 {\it Nucl. Phys. A} {\bf 617} 
        449; 
        Waas T et al. 1996 {\it Phys. Lett. B} {\bf 365} 12;
        1996 {\it Phys. Lett. B} {\bf 379};
        Sibirtsev A et al. 1998 {\it Nucl. Phys. A} {\bf 641} 476. 
\bibitem{Tsushima1}
        Tsushima K et al. 
        1998 {\it Phys. Lett. B} {\bf 429} 239; 
        (E) 1998 {\it ibid.} {\bf 436} 453.
\bibitem{Barth}
        Barth R et al. 1997 {\it Phys. Rev. Lett.} {\bf 78} 4007;
        Laue F et al., Phys. Rev. Lett. {\bf 82} 1640 (1999);
        Senger P and Strobele H 1999 {\it J. Phys. G} {\bf 25}
        R59.
\bibitem{Senger3}
        Senger P et al. 1999 {\it Prog. Part. Nucl. Phys.} {\bf 42}
        209.         
\bibitem{Li2}
        Li G Q, Lee C H and Brown G E 1997 {\it Nucl. Phys. A} {\bf 625}
        372;
        Cassing W and  Bratkovskaya E L 1999 {\it Phys. Rep.} {\bf 308}
        65;
        Li G Q, Ko C M and Chung W S 1998 {\it Phys. Rev. C} {\bf 57}
        434. 
\bibitem{Bratkovskaya}
        Bratkovskaya E L, Cassing W and Mosel U 1997
        {\it Nucl. Phys. A} {\bf 622} 593.

\bibitem{Schroter}
        Schroter A et al. 1994 {\it Z. Phys. A} {\bf 350} 101;
        Senger P et al. 1996 {\it Acta Phys. Polon. B} {\bf 27} 2993.
\bibitem{Guichon} 
        Guichon P A M 1989 {\it Phys. Lett. B} {\bf 200} 235. 
\bibitem{Guichonf}
        Guichon P A M et al. 
        1996 {\it Nucl. Phys. A} {\bf 601} 349;
        Saito K, Tsushima K, Thomas A W 1996 
        {\it Nucl. Phys. A} {\bf 609} 339.
\bibitem{Tsushima2}
        Tsushima K et al.
        1999 {\it Phys. Rev. C} {\bf 59} 2824; 
        Sibirtsev A et al. 1999 {\it Eur. Phys. J. A} 
        {\bf 6} 351; 
        Tsushima K et al. 
        1998 {\it Phys. Lett. B} {\bf 443} 26; 
        Tsushima K 2000 {\it Nucl. Phys.} {\bf A670} 198c; 
        Saito K et al. 1997
        {\it Phys. Rev. C} {\bf 55} 2637;
        Tsushima K et al. 1998 
        {\it Nucl. Phys. A} {\bf 630} 691.
\bibitem{Tsushima3}
        Tsushima K, Huang S W and Faessler A 1994 
        {\it Phys. Lett. B} {\bf 337} 245; 
        1995 {\it J. Phys. G} {\bf 21} 33;  
        1997 {\it Austral. J. Phys.} {\bf 50} 35; 
        Tsushima K, Sibirtsev A and  Thomas A W 1997 
        {\it Phys. Lett. B} {\bf 390} 29;  
        Tsushima K et al. 1999
        {\it Phys. Rev. C} {\bf 59} 369; (E) 2000 {\it ibid.} {\bf 61} 029903; 
        Sibirtsev A, Tsushima K and Thomas A W 1998
        {\it Phys. Lett. B} {\bf 421} 59; 
        Sibirtsev A et al. 
        1999 {\it Nucl. Phys. A} {\bf 646} 427.
\bibitem{LB}
        Landolt-B\"ornstein, New Series, ed. Schopper H 1973 {\bf 8}. 
\bibitem{Hombach}
        Hombach A et al. 1999 {\it Eur. Phys. J. A}
        {\bf 5} 77.
\bibitem{CBJ}
        Cassing W, Bratkovskaya E L and Juchem S {\it Nucl. Phys. A}
        in press.
\end{thebibliography}
\end{document}